\documentclass[aps,preprint,preprintnumbers,nofootinbib,showpacs]{revtex4-1}

\usepackage{amssymb,graphics,graphicx,color,epstopdf,subfigure,dcolumn,bm,slashed,tabularx}

\def\roughly#1{\mathrel{\raise.3ex\hbox
{$#1$\kern-.75em\lower1ex\hbox{$\sim$}}}}

\begin{document}

\title{Multi high charged scalars in the LHC searches and Majorana neutrino mass generations }

\author{Chian-Shu~Chen$^{1}$\footnote{chianshu@phys.sinica.edu.tw}, Chao-Qiang~Geng$^{1,2,3}$\footnote{geng@phys.nthu.edu.tw}, 
Da~Huang$^{2}$\footnote{dahuang@phys.nthu.edu.tw}
and
Lu-Hsing Tsai$^{2}$\footnote{lhtsai@phys.nthu.edu.tw}}
  \affiliation{$^{1}$Physics Division, National Center for Theoretical Sciences, Hsinchu, Taiwan 300\\
$^{2}$Department of Physics, National Tsing Hua University, Hsinchu, Taiwan 300\\
$^{3}$College of Mathematics \& Physics, Chongqing University of Posts \& Telecommunications, Chongqing, 400065, China}

\date{Draft \today}
\begin{abstract}
One natural way to understand
the excess of the measured $H\to\gamma\gamma$ rate 
over the standard model (SM) expectation
at the Large Hadron Collider (LHC) is to have  charged scalar bosons, existing  in  most of the SM extensions.
Motivated by this LHC result, we explore if it also sheds light on solving
the small neutrino mass generation problem.
We concentrate on a class of models with high dimensional  representations of scalars
to realize Majorana neutrino masses at two-loop level without imposing any new symmetry.
In these models, multi scalars with the electric charges higher than two are naturally expected,
which not only enhance the $H\to\gamma\gamma$ rate, but provide more searching grounds at the LHC.
In particular, the rate of $H\to Z\gamma$ also changes similar to that of the diphoton channel.
\end{abstract}

\maketitle

{\it Introduction}---
The Higgs doublet ($H$) not only breaks the electroweak gauge
symmetries $SU(2)_{L}\times U(1)_{Y}$ to $U(1)_{Q}$
but also gives masses to the charged fermions in the standard model (SM).
The main goals of the
Large Hadron Collider (LHC) are the search of the Higgs boson ($H$) and the investigation of  the electroweak symmetry
breaking mechanism. Recently, a boson with its mass around  $125$ GeV has been discovered by both ALTAS~\cite{:2012gk} 
and CMS~\cite{:2012gu} collaborations, 
and its spin-parity property is further identified to be $0^{+}$ 
based on the LHC full 2011+2012 dataset~\cite{moriond1}. 
The scalar is most likely the Higgs particle as its properties 
are consistent with the SM predictions except the possible large production rate of  $H\to\gamma\gamma$. 
In 2012, the excess of the rate in  both experiments is around $2\sigma$ deviation from the SM prediction. 
Currently, the measured signal 
strength of Higgs to diphoton by the ATLAS collaboration is $\mu = 1.6^{+0.3}_{-0.3}$~\cite{ATLAS_NOTE_2013_034},
 whereas the corresponding one by  the CMS collaboration
goes down to $\mu = 0.78^{+0.28}_{-0.26}$~\cite{moriond2}. In other words, the excess still survives in the ATLAS data
but disappears in the CMS measurement. 
The elimination of the inconsistency between the two collaborations will rely on the future data accumulation and analysis in the LHC 
phase-II. In this study, we consider the possibility that some of the excess is sustained. Clearly, in such a case, it is still a call for new physics.
One of natural mechanisms is to include new charged particles in the SM~\cite{Batell:2011pz,Carena:2012xa}, 
which would enhance the decay rate due to the new charged loop contributions.

On the other hand, neutrino oscillations observed by the solar, atmospheric, and reactor
neutrino experiments reveal  that neutrinos are massive but tiny and mix with each other, 
implying the need
of new physics as there is no renormalizable Yakawa interaction for neutrinos related to $H$ in the SM. 
Theoretical studies on model buildings to understand the small neutrino masses 
are enormously varied if it is allowed to freely invent
new particles, scalars and/or fermions, beyond the SM ones.
Without a new fermion, the only possibility of having non-zero neutrino masses is that neutrinos are
Majorana fermions with  the accidental global lepton symmetry in the SM  either explicitly or spontaneously broken.
Apart from the  seesaw mechanism~\cite{Minkowski:1977sc},
models with tiny Majorana neutrino masses arising from quantum corrections have been proposed~\cite{Zee:1980ai,Zee:1985id}.
It is clear that this type of theories is usually incorporated with
 some extension of the scalar sector in the SM. In particular, singly and doubly charged scalars are required in the
radiative neutrino mass generation mechanisms at one and two-loop levels, respectively.
As a result, the existence of new charged particles is a generic feature of the radiative neutrino models.
These charged scalars in turn would help for resolving the excess of the $H\to\gamma\gamma$ rate at the LHC
as discussed in the literature~\cite{Akeroyd:2012ms}. 
However, in most of the above radiative models, since only $SU(2)_L$ singlet scalars
are introduced, new physics effects are limited in the lepton sector, whereas those involving hadrons,
such as the neutrinoless double beta decays believed as a benchmark of the Majorana nature of neutrinos,
do not show up.

In Ref.~\cite{Chen:2006vn}, an unconventional neutrino mass generation model with a triplet scalar was proposed,
leading to two doubly charged scalars with the neutrinoless double beta decays dominated by the short distance contribution.
However, this type of models needs adding some {\em ad hoc} discrete symmetry
to forbid the neutrino mass term at tree level~\cite{Chen:2010ir},
which makes these models less appealing.
In this study, we examine the most general cases with
arbitrary $SU(2)_L$ representations for scalars to avoid the tree level
neutrino mass generation without imposing any new symmetry.
It turns out that a class of models with high odd-dimensional scalars, which generate neutrino masses at two-loop level,
exists naturally.
Consequently, multi scalar bosons with the electric charges higher than 2-unit appear.
These multi charged scalars clearly ensure us some rich phenomenologies at the LHC. 
We will explore them along with the lepton number violating processes.
 
{\it Multi scalars}---
For all non-Higgs like scalars with non-trivial $SU(2)_L\times U(1)_Y$ quantum numbers~\cite{nocolor},
there are only three possible renormalizable Yukawa interactions, given by
\begin{eqnarray}\label{yukawa}
f_{ab}\bar{L}^{\rm c}_{a}L_{b}s \;,   \quad y_{ab}\bar{\ell}^{\rm c}_{R_a}\ell_{R_{b}}\Phi\;, \quad  {\rm and} \quad
g_{ab}\bar{L}^{\rm c}_{a}L_{b}T\;,
\end{eqnarray}
where $L$ $(\ell_{R})$ stands for the left-handed (right-handed) lepton, $a$ or  $b$ denotes $e, \mu$ and $ \tau$,  c represents the charged
conjugation,   $s$ and $\Phi$ are $SU(2)_{L}$
singlet scalar fields with $Y = 2$ and $Y = 4$, and $T$ is an $SU(2)_{L}$ triplet with the hypercharge $Y = 2$,
respectively~\cite{convention}.
Note that the triplet scalar $T$ can
 generate neutrino masses at tree level after developing its vacuum expectation value (VEV), known as
the Type-II seesaw mechanism~\cite{Magg:1980ut}, while $s$ and $\Phi$ are used 
in  Zee~\cite{Zee:1980ai} and Zee-Babu~\cite{Zee:1985id} models, respectively.

In Eq.~(\ref{yukawa}), the third Yukawa interaction is the most troublesome as it generates neutrino masses at tree level with an extreme small
value of the VEV or Yukawa couplings, which is obviously un-natural. Without introducing the triplet $T$, the interactions in Eq.~(\ref{yukawa})
are precisely given by the Zee-Babu model~\cite{Zee:1985id}, which has been extensively studied in the literature, in particular its
phenomenology of the doubly charged scalar at the LHC.
In this work, we consider a new class of models by adding
a scalar field $\xi$ with $Y = 2$ and a non-trivial  $SU(2)_L$
representation ${\bf n}$.
To minimize our models, we disregard the singlet scalar $s$ and keep the other singlet
 $\Phi$ so that only the second Yukawa
interaction in Eq.~(\ref{yukawa}) can exist at tree level.

The general scalar potential reads
\begin{eqnarray}\label{potential1}
V(H, \xi, \Phi^{\pm\pm}) &=& -\mu_{H}^2|H|^2 + \lambda_{H}|H|^4 + \mu^2_{\xi}|\xi|^2 + \lambda^{\alpha}_{\xi}|\xi|_{\alpha}^4 \nonumber \\
&+& \mu^2_{\Phi}|\Phi|^2 + \lambda_{\Phi}|\Phi|^4 + \lambda^{\beta}_{H\xi}(|H|^2|\xi|^2)_{\beta} \nonumber \\
&+&  \lambda_{H\Phi}|H|^2|\Phi|^2 + \lambda_{\xi\Phi}|\xi|^2|\Phi|^2
+ [ \mu\xi\xi\Phi + \rm h.c.],
\end{eqnarray}
where $\alpha$, $\beta$ are the short-handed notations denoting the possible invariant terms  for 
higher representations in general. Also notice that all terms in the potential are self-hermitian except the last $\mu$-term,
which is related to the dynamics of the lepton number breaking to be discussed later.
For the even dimensional representations, {\em i.e.} ${\bf n}= {\bf 2, 4, 6}\cdots$, the products $\xi\xi$
vanish since
\begin{eqnarray}
\xi\xi &=& \epsilon_{ii'}\epsilon_{jj'}\epsilon_{kk'}\cdots\xi_{ijk...}\xi_{i'j'k'...} \nonumber \\
&=& -\epsilon_{i'i}\epsilon_{j'j}\epsilon_{k'k}\cdots\xi_{ijk...}\xi_{i'j'k'...} = 0\,,
\end{eqnarray}
due to  the anti-symmetric matrix of $\epsilon_{ij}$ ($i,j$ = 1,2).
Subsequently, we only need to consider
the odd dimensional representations of $\xi$, {\em i.e.} ${\bf n}= {\bf 3, 5, 7}\cdots$. Since the triplet has been dropped out,
the next minimal choice is ${\bf n}= {\bf 5}$.
From now on, we concentrate on this minimal one,  {\em the quintuplet}, with
$\xi = (\xi^{+++}, \xi^{++}, \xi^{+}, \xi^{0}, \xi^{-})^T$.
One shall bear in mind that the results can be easily extended to those with higher representations of
$\xi$~\cite{RGeffect,Cirelli:2005uq}. In these cases, there are three and two irreducible terms for $|\xi|^4$ and $|H|^2|\xi|^2$ respectively.
 
We now move to the lepton number $(L)$ violation. In general, $U(1)_{L}$ can be either global
 or gauge symmetry.
 If the lepton number indeed comes from a global symmetry as that in the SM, the
spontaneous symmetry breaking will generate a Nambu-Goldstone (NG) boson, usually called Majoron.
In this model, the VEV of $v_\xi/\sqrt{2}=\langle \xi^0 \rangle$
breaks  $U(1)_L$ spontaneously~\cite{rhoparameter}. 
Since $\xi$ is an $SU(2)$
multiplet, its corresponding Majoron has a direct coupling to the $Z$ boson, which is strongly  constrained  by the LEP
measurement of the invisible $Z$ decay width.
To resurrect it, we illustrate here by
adding another $SU(2)_{L}$ triplet ($\bf 3$) scalar field $\Delta=(\Delta^{+}, \Delta^{0}, \Delta^{-})^T$ with $Y = 0$
to the model, which results in one additional non-hermitian term $\Delta\xi H^*H^*$.  
It is easy to see that the coexistences of  $\bar{\ell}^c_{R}\ell_{R}\Phi$ , $\mu\xi\xi\Phi$ and
$\Delta\xi H^*H^*$ break the lepton number explicitly. 
Since  $v_\xi$ is constrained to 
be less than around $1~\mathrm{GeV}$ by the $\rho$ parameter~\cite{rhoparameter,pdg}, it can be shown that the mixing angle between the 
neutral scalar components of $\xi$ and $H$ is of  ${\cal O}(v_\xi/v)$. In practice, it is safe to ignore the mixing and 
treat the neutral component from the doublet field $H$ as the SM-like Higgs. In addition, as  $\xi$ does not couple to the SM fermions 
directly,  the single production of $\xi^0$ via the gluon-gluon fusion process as well as  its $W^+W^-$ fusion and W-associated productions
are all suppressed by a factor of $(v_{\xi}/v)^2 \leqslant 10^{-4}$. 

\begin{figure}[t]
  \centering
  \includegraphics[width=0.45\textwidth]{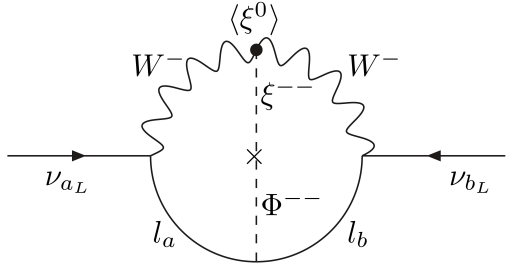}
  \caption{Two-loop contributions to neutrino masses.}\label{Fig_numass}
\end{figure}

\begin{figure}[t]
  \centering
  \includegraphics[width=0.45\textwidth]{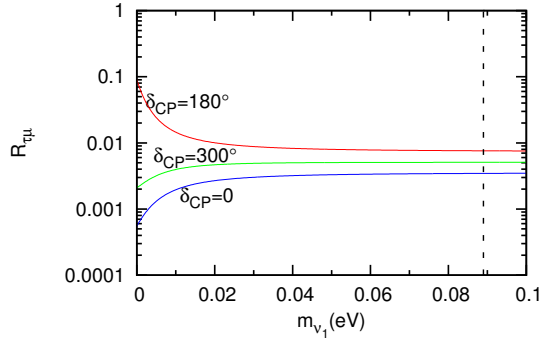}
  \caption{$R_{\tau\mu}$ versus $m_{\nu_1}$, where the blue, red and green curves correspond to
  $\delta_{CP} = 0$, $180^\circ$ and $300^\circ$, respectively, with
the global  $\chi^2$ analysis in Ref.~\cite{GonzalezGarcia:2012sz}, while the dashed black line represents the
  lower bound set by cosmology.}\label{Fig_mn1ratio}
\end{figure}

{\it Neutrino masses}---
In the Zee-Babu model, neutrino masses are generated from
the two-loop diagrams due to the couplings of $s^{\pm}s^{\pm}\Phi^{\mp\mp}$.
whereas in our model without $s^{\pm}$,
they can be induced from similar two-loop diagrams
with $\Phi^{\pm\pm}$ coupling to $W^{\mp}W^{\mp}$ though the mixing of $\xi$
as shown in Fig.~\ref{Fig_numass}.
In this mechanism, neutrino masses are calculable, given by
\begin{eqnarray}
m_{\nu_{ab}} &\simeq& \frac{g^4}{\sqrt{2}(4\pi)^4}m_{a}m_{b}v_{\xi}y_{ab}\sin{2\theta} \nonumber \\
&\times& \left[\frac{1}{M^2_{P_1}}\log^2{\left(\frac{M^2_{W}}{M^2_{P_1}}\right)} - \frac{1}{M^2_{P_2}}\log^2{\left(\frac{M^2_{W}}{M^2_{P_2}}\right)}\right],
\end{eqnarray}
where $m_{a,b}$ correspond to charged lepton masses; $P_{1}$ and $P_{2}$ are the mass eigenstates of doubly charged scalars with
$\theta$ representing their mixing angle and  $M_{P_i} > M_{W}$  assumed.
Note that the neutrino masses are suppressed by the two-loop
factor, $SU(2)_L$ gauge coupling, charged lepton masses, mixing angle $\theta$, and VEV of $\xi$, respectively, without fine-tuning
Yukawa couplings $y_{ab}$.
The model predicts the neutrino mass spectrum to be a normal hierarchy if one requires the
perturbative bound on $y_{ab}$. Consequently, the neutrino mass matrix is given by
\begin{eqnarray}
m_{\nu_{ab}} = U_{\rm PMNS}m_{\nu_{diag}}U^{T}_{\rm PMNS} \propto y_{ab}\,,
\end{eqnarray}
where
$m_{\nu_{diag}}$ = diag($m_{\nu_1}$, $\sqrt{m^2_{\nu_1} + \Delta m^2_{sol}}$, $\sqrt{m^2_{\nu_1} + \Delta m^2_{atm}}$)
with $m_{\nu_1}$ the lightest $\nu$ mass.
 From the neutrino oscillation data, one is able to pin down the neutrino parameters
 via the leptonic processes governed by $y_{ab}$. For example,
 the ratio $R_{\tau\mu} \equiv \frac{Br(\tau \rightarrow e\gamma)^*}{Br(\mu \rightarrow e\gamma)}$~\cite{BR}
is related to the lightest neutrino mass, $m_{\nu_1}$, as illustrated in
Fig.~\ref{Fig_mn1ratio}, with the use of the latest neutrino oscillation data~\cite{GonzalezGarcia:2012sz}. 
In principle,  many variants of this quantity can be also defined from the leptonic rare decays or
same-sign dilepton decays of $\Phi^{\pm\pm}$. 
Hence, our model provides a complementary way to 
determine neutrino parameters.
Due to the indirect couplings between $\Phi^{\pm\pm}$ and $W^{\mp}W^{\mp}$,  
our model reveals
a mechanism for possible large neutrinoless double $\beta$ decays which do not
directly involve the small Majorana neutrino masses unlike the Zee-Babu model.

\begin{figure}[t]
  \centering
  \includegraphics[width=0.35\textwidth]{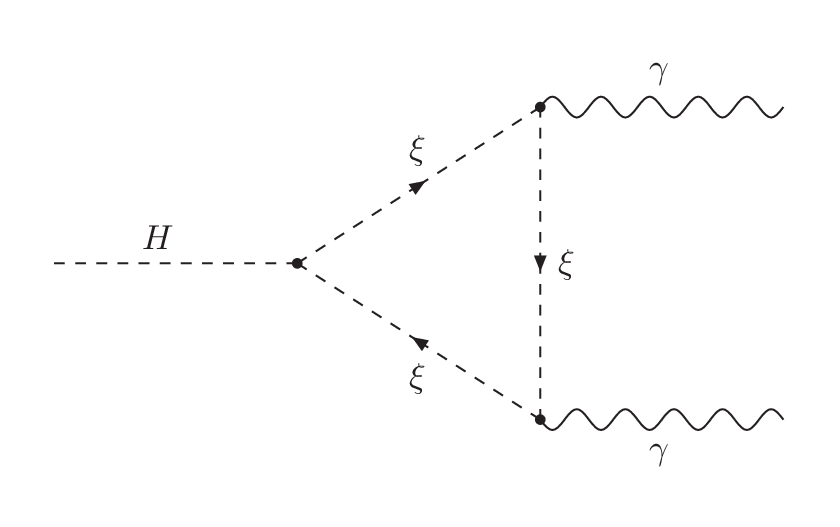}
   \includegraphics[width=0.35\textwidth]{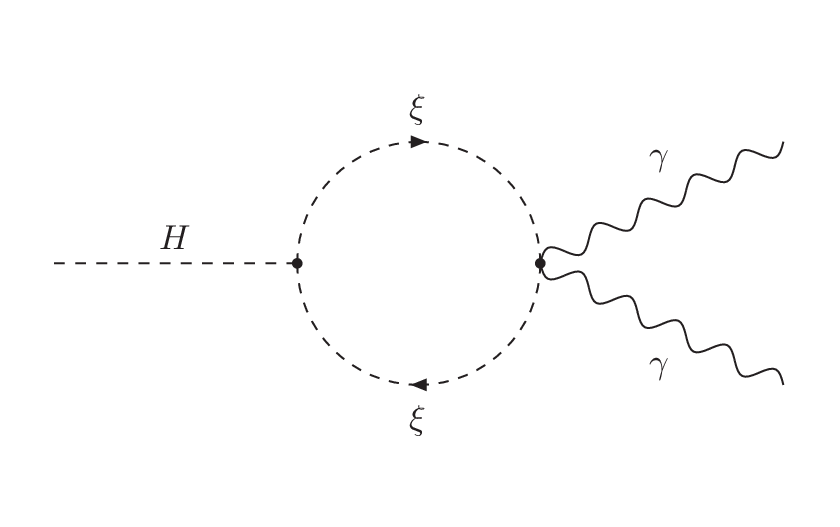}
  \caption{Contributions to $H\rightarrow \gamma\gamma$ from charged scalar exchanges in the loops.
  }\label{Fig_diphotondiagram}
\end{figure}

\begin{figure}[t]
  \centering
  \includegraphics[width=0.45\textwidth]{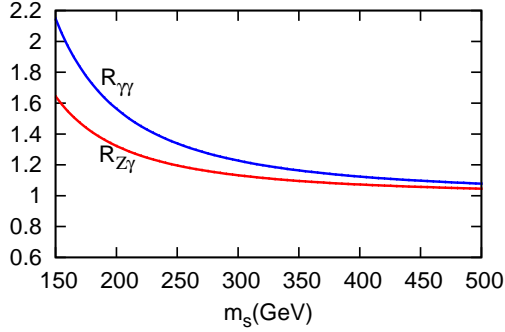}
  \caption{$R_{\gamma\gamma}\equiv \Gamma(H\to\gamma\gamma)/\Gamma(H\to\gamma\gamma)_\mathrm{SM}$ and 
  $R_{Z\gamma} \equiv \Gamma(H\to Z\gamma)/\Gamma(H\to Z\gamma)_\mathrm{SM}$ as functions of the degenerate 
  mass factor $m_s$ of the multi charged scalar states with ${\bf n=5}$ and the universal trilinear coupling to Higgs, 
  $\mu_{s}$ = -100 GeV.
  }\label{Fig_Higgstwophoton}
\end{figure}

{\it LHC searches}---
In our minimal model with ${\bf n=5}$,
apart from one SM-like Higgs scalar with $m_H^2=2\lambda_Hv^2$, there are one pseudo-scalar, three singly
charged, two doubly charged, and one triply charged scalars with masses  all around a few hundred GeV,
  which all satisfy the current experimental bounds~\cite{pdg}.
Since $\xi$ does not directly interact with the SM fermions, the Higgs production is not expected to be modified.
However, as promised,
the decay rate of $H\to\gamma\gamma$ receives extra contributions from the new charged scalars in the loops
as shown in Fig.~\ref{Fig_diphotondiagram}, given by
\begin{eqnarray}
\Gamma(H \rightarrow \gamma\gamma)&=&{G_F\alpha^2 m_H^3\over 128\sqrt{2}\pi^3}|\sum_{f}N^c_fQ_f^2A_{1\over 2}(\tau_f)+A_1(\tau_W) \nonumber \\
&+& \sum_{I_{3}}(I_{3} + 1)^2{v\over2}{\mu_{s}\over m_{s}^2}A_0(\tau_{s})|^2\,,
\end{eqnarray}
where the third term corresponds to those from the multi  charged components of the dimension-${\bf n}$ scalar multiplet $\xi$ with
  their electric charges $(I_{3} + 1)$,
  $I_{3}$ runs from $-(\mathbf{n}-1)/2$ to $(\mathbf{n}-1)/2$, and $\mu_{s}$ is the trilinear
coupling to the SM Higgs. The amplitudes $A_{0,{1\over 2},1}$
and the mass ratios $\tau_{f,W,s}$ are defined
in Ref.~\cite{Djouadi:2005gi}. Here, for simplicity,
we have taken the same trilinear coupling $\mu_s$ and charged scalar mass $m_s$.
The new contributions from the multi charged scalars interfere constructively with that of the SM if $\mu_s<0$.
As an illustration,
we plot the ratio of $R_{\gamma\gamma}\equiv \Gamma(H\to\gamma\gamma)/\Gamma(H\to\gamma\gamma)_{SM}$
in Fig.~\ref{Fig_Higgstwophoton}
with a typical value of $\mu_{s} = -100$ GeV.
It is clear that the excess for  the diphoton decay rate reported by  ATLAS 
can be easily explained by these multi charged scalars. Furthermore, the partial width of $H \rightarrow Z\gamma$ also 
receives new contributions from the multi charged scalar fields in our model.
In  Fig.~~\ref{Fig_Higgstwophoton}, we plot 
$R_{Z\gamma}\equiv \Gamma(H\to Z\gamma)/\Gamma(H\to Z\gamma)_{SM}$ with ${\bf n=5}$.
Like the diphoton channel,
$\Gamma(H \rightarrow Z\gamma)$ also gets enhanced~\cite{Zgamma} due to the new charged fields
in comparison with
the SM prediction. 
The simultaneous observations of $\gamma\gamma$ 
and $Z\gamma$ modes would help for discriminating the possible new physics beyond the SM~\cite{Carena:2012xa}.

The scalar multiplet components can be pair produced via $q\bar{q} \rightarrow \gamma/Z^* \rightarrow \xi\xi^* (\Phi^{++}\Phi^{--})$
or  $q\bar{q}' \rightarrow W^* \rightarrow \xi^{+++}\xi^{--} (\xi^{++}\xi^{-},\xi^{+}\xi^0)$ at the LHC. Similar processes
for the triplet Higgs productions
have been investigated in the Type-II seesaw mechanism~\cite{Akeroyd:2005gt}. 
The $\xi^{++}\xi^{--}$-pair production rate 
varies with $m_{\xi}$ from 1 to $10^{-3}$ pb for 
$m_{\xi} = 100$ to 600~GeV with $\sqrt{s} = 8$ TeV.
The lower mass limit of $\xi^{\pm\pm}$ has been set by  CMS~\cite{Chatrchyan:2012ya} and 
ATLAS~\cite{ATLAS:2012hi} to be
(382, 391, 395) and (409, 375, 398) GeV by assuming 
100\% branching ratio  of ($e^{\pm}e^{\pm}$, $e^{\pm}\mu^{\pm}$, $\mu^{\pm}\mu^{\pm}$) final states, respectively. 
The bounds, of course, will reduce significantly if more leptonic modes are opened.
One noticeable feature of our model is 
that there are two doubly charged states, $\xi^{\pm\pm}$ and $\Phi^{\pm\pm}$, 
which  decay almost 100\% into 
$W^{\pm}W^{\pm}$ and $l^{\pm}_{R}l^{\pm}_{R}$, respectively. In particular, the channels of $\xi^{\pm\pm}\to \xi^{\pm}\xi^{\pm}$ 
are kinematically forbidden.
Therefore, the previous experimental bounds can be directly applied 
to $\Phi^{\pm\pm}$. 
For the $\xi^{++}\xi^{--}$-pair productions,
in Table~\ref{SB} 
we list the cross sections of the signals with $m_{\xi} = 100 - 500$~GeV 
and the corresponding SM backgrounds, where we only consider the final lepton state to be 
$e$ or $\mu$ due to the low tagging efficiency of $\tau$.
\begin{table}[t]
\begin{tabular}{|c|c|c|}\hline
$m_{\xi}$ (GeV) & Signal [fb] & Signal [fb]  \\\hline\hline
 & $2\ell^{+}2\ell^{-}~+\slashed{E}_{T}$  & $\ell^{+}2l^{-}2j~+\slashed{E}_{T}$ \\
100 & $5.2\times10^{-1}$ & $3.3$  \\
200 & $8.2\times10^{-2}$ & $5.2\times 10^{-1}$  \\
300 & $2.1\times10^{-2}$ & $1.3\times10^{-1}$ \\
400 & $6.7\times10^{-3}$ & $4.2\times10^{-2}$ \\
500 & $2.5\times10^{-3}$ & $1.6\times10^{-2}$ \\\hline
 & Backgrounds [fb] & Backgrounds [fb] \\    
 & $2W^{+}2W^{-}$ ~~~ $W^{+}W^{-}Z$ & ~~~~~~~~~$t\bar{t}W^{-}$~~~~~~~$2W^{-}W^{+}2j$~~~~~~~~$W^{-}Z2j$ \\
 & $2.2\times10^{-4}$~~~~~~~$6.6\times10^{-2}$ & $3.6\times10^{-1}$~~~~~$6.7\times10^{-2}$~~~~~~~~~~$17.8$  \\\hline
\end{tabular}
\caption{\label{SB} Cross sections of the $\xi^{++}\xi^{--}$-pair productions and the corresponding SM backgrounds with $\sqrt{s} = 8$~TeV,
 where  $\ell=e$ or $\mu$.}
\end{table}
 From the table, we see that there is only less than one signal event based on the current 
luminosity of about $20~\rm{fb^{-1}}$ at the LHC if $m_{\xi}$ is heavier than 200~GeV  for four charged lepton modes. 
For the signals with more jets, we expect higher SM backgrounds although the kinematical cuts are usually used to enhance the signal significance. 
A detailed collider analysis is beyond the scope of this brief report. Furthermore, if we take  
$R_{\gamma\gamma} \simeq 1.2$ of the combined result from ATLAS and CMS collaborations, the $m_{\xi}$ is about 300~GeV.  

The signals of the triply charged component in the quintuplet scalar are distinct
and its production rate is about
two times larger than the doubly charged one according to their coupling strengths.
The triply charged scalar $\xi^{+++}$
decays into three W bosons with the width
\begin{eqnarray}
\Gamma(\xi^{+++}\rightarrow 3W^+)\simeq{3g_W^6v_\xi^2 m_\xi^5\over 512\pi^3m_W^6}\,,
\end{eqnarray}
which subsequently go into the final states with
six fermions,
where we have ignored the phase space suppression.
Note that in the degenerate limit, all W bosons can be on-shell.
In addition to the multiplet $\xi$, the doubly charged scalars $\Phi^{\pm\pm}$
also make the signals of our model more distinct at the LHC. If $m_{\xi^{+++}} > m_{\Phi^{++}}$,
the channel $\xi^{+++} \rightarrow \Phi^{++} + W^{*+}$ opens up for the triply charged scalar
through the mixings,
which can be distinguished from the $3W^{*+}$ one as
$\Phi^{\pm\pm}$ only decay to the same sign right-handed charged leptons with large center masses.
By tagging the taus produced
from $\Phi$ or  $W$  one can pin down the chirality of $\tau$ from its subsequent pion distribution
spectrum~\cite{Tsai:1971vv}.
Finally, we remark that the above results can be easily applied to
other models with higher charged scalars.

{\it Summary}---
We have proposed a common origin  for  the excess of the  $H\to\gamma\gamma$ rate measured by the LHC
and small neutrino masses based on models with some multi high charged scalars.
We have shown that due to the charged scalars, 
 both rates of $H\to\gamma\gamma$ and $H\to Z\gamma$ can be enhanced,
while small Majorana neutrino masses are realized radiatively through the two-loop diagrams.
Our models predict a normal hierarchy for the neutrino mass spectrum  
and
allow a large rate for the neutrinoless double $\beta$ decay.
We have demonstrated that the multi high charged scalars
 provide many interesting searching grounds at the LHC.

\subsection*{Acknowledgement}
This work was supported in part by National Center for Theoretical Sciences,
National Tsing-Hua University (101N1087E1) and National Science Council
(NSC-98-2112-M-007-008-MY3 and NSC-101-2112-M- 007-006-MY3), Taiwan, R.O.C.


\begin{thebibliography}{99}
\bibitem{nocolor}
We assume that scalars carry no color.
\bibitem{rhoparameter}
For $\xi$ being a
triplet or higher representation,
it will contribute to  the $\rho$ parameter.
The experimental value of $\rho = 1.0004^{+0.0003}_{-0.0004}$~\cite{pdg}
leads to a constraint of
$\langle \xi^0 \rangle \leqslant {\cal O}(1)$ GeV.
\end{thebibliography}
\end{document}